\documentclass[a4paper]{aa}
\def \src {XB\thinspace 1323$-$619 }
\def \th {\thinspace}
\def \xte {{\it Rossi-XTE }}
\def \degmark{^\circ}

\def \hcm {\hbox {\ifmmode $ atoms cm$^{-2}\else atoms cm$^{-2}$\fi}}
\def \arcmin {\hbox{$^\prime$}}
\def \arcsec {\hbox{$^{\prime\prime}$}}

\def \approxgt{\mathrel{\hbox{\rlap{\lower.55ex \hbox {$\sim$}}
        \kern-.3em \raise.4ex \hbox{$>$}}}}
\def \approxlt{\mathrel{\hbox{\rlap{\lower.55ex \hbox {$\sim$}}
        \kern-.3em \raise.4ex \hbox{$<$}}}} 
\usepackage{graphicx}
\begin{document}

%   \thesaurus{6(13.25.5; % X-rays: stars,
%               08.09.2: X\th 1624$-$490;  % stars: individual: \src,
%               08.14.1;  % stars: neutron,
%               08.02.1;  % binaries: close,
%               02.01.2;  % accretion, accretion disks
%               09.04.1)} % ISM: dust, extinction
%
\title{A study of the spectral evolution during dipping in XB\th
1323-619 with Rossi-XTE and BeppoSAX}
\author{R. Barnard\inst{1}
        \and M. Ba\l uci\'nska-Church\inst{1,2}
        \and A.P. Smale\inst{3}
        \and M.J. Church\inst{1,2}}
\offprints{R. Barnard}
\institute{School of Physics and Astronomy, University of Birmingham,
              Birmingham B15 2TT\\
              email: robb@star.sr.bham.ac.uk, mbc@star.sr.bham.ac.uk, mjc@star.sr.bham.ac.uk
\and
              Astronomical Observatory, Jagiellonian University, ul. Orla 171,
              30-244 Cracow, Poland
\and
              Laboratory for High Energy Astrophysics, Code 662,\\
              NASA/Goddard Space Flight Center, Greenbelt, MD~20771, USA\\
              email: alan@osiris.gsfc.nasa.gov}
%   \thanks{}
\date{Received 18 May 2001; Accepted 23 September 2001}
\authorrunning{R. Barnard et al.}
\titlerunning{A study of \src during dipping}
\abstract
{We report results from analysis of the observations of the dipping low mass X-ray 
binary  XB\thinspace 1323$-$619 made with {\it BeppoSAX} and {\it
Rossi-XTE}. The dust-scattered halo contributes significantly in this source, and
the observation made with {\it BeppoSAX} on 1997 August was
used to provide MECS radial intensity profiles at several energies.
From these, the halo fractions were obtained and thus an optical depth to
dust scattering of 1.8$\pm$0.4 derived. In the \xte observation of 
April 25--28, 1997, seven X-rays dips were observed together with 7 bursts
repeating approximately periodically. Non-dip and dip PCA spectra can be
well-described by assuming the emission consists of point-like
blackbody emission identified with the neutron star, plus Comptonized
emission from an extended ADC. The blackbody temperature is 1.79$\pm$0.21 keV 
and the cut-off power law photon index 1.61$\pm$0.04. Spectral evolution in 
dipping is well described by progressive covering of the extended Comptonizing 
region by absorber plus more rapid removal of the point-like blackbody. 
The effects of dust scattering and of the X-ray pulsar 1SAX\th J1324.4-6200 
also in the field of view are included in the fitting. We detect 
an iron line at $\sim$6.4 keV and its probable origin in the ADC is
discussed.
\keywords   {X rays: stars --
             stars: individual: \src\ --
             stars: neutron --
             binaries: close --
             accretion: accretion disks --
             ISM: dust, extinction}}
\maketitle 

\section{Introduction}

\src is a member of the dipping class of Low Mass X-ray Binaries (LMXB) that exhibit
irregular reductions, or dips, in X-ray intensity at the orbital period. It is 
generally accepted that these are caused by absorption in the bulge in the outer 
accretion disk where the flow from the companion star meets the disk (White \& 
Swank 1982). XB\th 1323-619 is faint ($\sim $3 mCrab), has a period of 2.94 hr 
and is remarkable as one of the small group of sources in which quasi-periodic 
bursting is observed. Investigation of spectral evolution during
dipping reveals not only the structure and properties of the outer
accretion disk, but also the nature and geometry of the emission
regions, since the requirement to fit non-dip and several dip spectra
strongly constrains emission models. In the dipping LMXB, the spectral
changes during dipping cannot be described by absorption of
one-component emission but require two emission components: point-like
blackbody emission identified with the surface of the neutron star plus
Comptonized emission from an extended Accretion Disk Corona (ADC)
(e.g. Church et al. 1997). During dipping, the Comptonized emission is removed 
gradually as the extended absorber progressively overlaps the ADC to 
increasing extents, whereas the blackbody emission is rapidly removed as the
absorber overlaps the point source. This model has been able to
explain spectral evolution during dipping in the dipping LMXB 
(Church et al. 1997, 1998a,b; Ba\l uci\'nska-Church et al. 1999, 2000).  
The other major type of model applied to the dipping sources
comprises thermal emission from the accretion disk plus a Comptonized
component (Mitsuda et al. 1989; Yoshida et al. 1995). 
Our two-component model has
also been applied in a survey of the other classes of LMXB, i.e. the
Z-track and Atoll sources, using {\it ASCA} and provided good fits to
all sources in the survey spanning a luminosity range 
$\rm {3\times 10^{36}}$ erg $\rm {s^{-1}}$ to $\rm {5\times 10^{38}}$
erg $\rm {s^{-1}}$ (Church \& Ba\l uci\'nska-Church 2001). This survey
showed that a blackbody component was present in all sources; moreover,
this was unlikely to originate in the accretion disk
as the required values of inner radius were substantially less than
the neutron star radius in many cases. However, assuming the emission was from
an equatorial belt on the neutron star revealed an agreement between
the half-height of this region, and the half-height of the inner,
radiatively-supported disk (Church 2001), suggesting possible mechanisms
by which the emitting area is determined (Church et al. 2001).
These results provide further strong evidence that the blackbody 
emission in LMXB originates on the neutron star.

In the case of the dipping LMXB, the size of the extended Comptonizing
ADC can be measured {\it via} dip ingress times. Application of
this technique to several dipping sources (Church 2001) has revealed
that the ADC is very extended, with radius typically 50,000 km or 15\%
of the accretion disk radius. Moreover, the ADC is {\it thin} having small
height-to-radius ratio (Smale et al. 2001) since the absorber will not
extend to the very large vertical distance required to cover a 
spherical region of vertical height 50,000 km. The very extended, thin nature 
of the ADC has several important consequences. Firstly, these
measurements show that models in which Comptonization takes place
in a central region, e.g. a small spherical region close to the
neutron star must be incorrect. Secondly, the ADC covers all of the
X-ray emitting disk. This together with the high optical depths to electron 
scattering of the ADC (Church 2001) means that all thermal emission
from the disk will be Comptonized and no disk blackbody emission will
be able to reach an observer. Thus blackbody emission from LMXB will
originate on the neutron star, not on the disk. Finally, as the disk
out to a radius of typically 50,000 km acts a source of Comptonization
seed photons, the net spectrum of this part of the disk is easily shown
to be very soft with $kT$ between 0.001 and 0.1 keV for typical
luminosities. Thus a recent tendency to use the {\sc comptt} model
for Comptonization with $kT$ $\sim $1 keV for the seed photons 
(e.g. Guainazzi et al. 1998) is inconsistent with the measurements of 
ADC radius above, and more important, is inconsistent with the
assumption in this model that the Wien approximation is valid, e.g.
in the range 1--10 keV, and so will lead to errors in spectral fitting results.

\src itself was first detected by {\it Uhuru} (Forman et al. 1978) and {\it Ariel V} 
(Warwick et al. 1981) and dipping and bursting discovered using {\it Exosat} 
(van der Klis et al. 1985; Parmar et al. 1989). Dip spectra revealed a  
component that was not absorbed in dipping, and spectral evolution was
modelled by dividing the non-dip spectral form into two components, one
of which was absorbed, the other unabsorbed but having decreasing
normalization in dipping. This ``absorbed + unabsorbed'' approach was
applied to several similar sources (Parmar et al. 1986; Courvoisier et
al. 1986; Smale et al. 1992), however, the changing
normalization was difficult to justify physically. The ``progressive covering''
model in which the absorber progressively overlaps an extended ADC
is able to explain the unabsorbed component simply as the uncovered
emission, in all such sources (e.g. Church et al. 1997; Smale et al. 2001). The
point-like blackbody component is rapidly covered. During the {\it Exosat} observation,
bursts repeated every 5.30--5.43 hr, approximately every second orbit. 

A detailed study of the source was made with 
a 120-ks {\it BeppoSAX} observation (Ba\l uci\'nska-Church et al. 1999)
during which 12 intensity dips and 10 type I X-ray bursts took place.
From the dipping, an orbital period of 2.938$\pm $0.020 hr was derived. Bursting 
repeated regularly as seen in {\it Exosat}, but with a timescale of
2.40--2.57 hr, i.e. smaller than the orbital period, so that bursts
marched through the dips leading to several occurrences of bursts
during dips. It was demonstrated that the spectra of bursts in dips were
consistent with a reduction in intensity in a totally ionized
absorber, resulting from the ionization of all parts of the accretion
disk by the bursts (Ba\l uci\'nska-Church et al. 1999). The {\it BeppoSAX} 
broadband non-dip spectrum in the range 1--150 keV was fitted simultaneously 
with 3 MECS dip spectra, and the best fit was obtained using the point-source 
blackbody plus an extended Comptonization model referred to above.
The Comptonization cut-off energy $E_{\rm CO}$ was found to be
44$^{+5}_{-4}$ keV, indicating a relatively high mean electron temperature in the ADC of
at least 15 keV. Dipping was well-described by combining this emission model 
with the progressive covering description of absorption. Quasi periodic 
oscillations have been detected in the quiescent, dipping and bursting emission 
(Jonker et al. 1999) using the {\it Rossi-XTE} observation made by the present authors.

We present here results of a study of dust scattering in \src\ made 
with {\it BeppoSAX}, and a study of dipping made with {\it
Rossi-XTE}. The radial distribution of intensity in the {\it BepppoSAX} MECS 
instrument is used to provide the dust-scattering cross section in
several energy bands, and from this the optical depth to dust scattering 
at 1 keV is derived. The relatively high Galactic column density ($\rm 
{\sim 4\times 10^{22}}$ atom $\rm {cm^{-2}}$, 
Ba\l uci\'nska-Church et al. 1999) means that dust scattering will
affect the observed intensity of the source, by scattering both out
of, and into the beam, and by introducing a time delay due to the
scattering process. This delay results in non-dip photons reaching
the observer during a dip, adding a contribution to the intensity
in dipping. In the {\it Rossi-XTE} observation of \src\, all of the
halo is collected as the PCA is non-imaging with a field-of-view of
1$\degmark$ and it is not possible to exclude the halo by selection 
from the central image as in the MECS.

In the present work, we have made a detailed study of spectral evolution
in dipping, for the first time including the effects of dust scattering
and utilizing the much higher sensitivity of the {\it Rossi-XTE} PCA
to improve on the spectral fitting study made with {\it BeppoSAX}
(Ba\l uci\'nska-Church et al. 1999). The effects of the X-ray pulsar 
1SAX\th 1324.4-6200 included in the field-of-view of the source are
added to the modelling. We also detect  an iron
line at $\sim$6.4 keV in this source.

\section{Observations}

\subsection{{\it Rossi-XTE}}

XB\th 1323-619 was observed with {\it Rossi-XTE} (Bradt et al. 1993)
from 1997, April 25 22:02:56 to April 28 03:54:40, the observation
spanning 200 ks. Results presented here were 
obtained using the Proportional Counter Array (PCA) instrument
operated in Standard 2 mode with 16~s time resolution. 
The PCA consists of 5 non-imaging, coaligned Xe multiwire proportional counter 
units (PCUs) with a 1$\degmark \times 1\degmark$ field of view and a total collecting 
area of $\sim$6500 cm$^{\rm {-2}}$ (Jahoda et al. 1996). All 5 PCUs were
operating throughout 95\% of this observation, and for optimum count
statistics we use only this data in spectral analysis, also using only the 
top detector layers to minimise detector background.
The data were screened to have elevation above the Earth's limb
$>$ 10$\degmark$, and angular deviation of the pointing axis of the
telescope from the source $<$ 0.02$\degmark$. 
PCA background subtraction was carried out using the latest versions of the 
appropriate background models for faint sources: the ``faint17/faint240''
models generated by the {\it Rossi-XTE} PCA team. Source and
background spectra were compared, and data rejected above energies at
which these become equal. Tests showed that in this faint source,
spectral fitting results were not better constrained by use of HEXTE
data. Light curves and spectra  were deadtime corrected using the {\it
Rossi-XTE} standard analysis software {\it Ftools 5.0.1}. Systematic 
errors of 1\% were added to the spectra. 
The field of view of the PCA contained the 170~s period X-ray pulsar 1SAX\th
J1324.4-6200 which was discovered serendipitously during the {\it
BeppoSAX} observation of XB\th 1323-619 (Angelini et al. 1998).
It is located 17$\arcmin$ from the LMXB, and although weak (1--10 keV
luminosity $sim$ $\rm {1.1\times 10^{34}}$ erg s$^{\rm -1}$)
for its lower limit distance of 3.4 kpc,  makes a
non-zero contribution to the spectrum of XB\th 1323$-$619, especially in deep
dipping when the count rate of the LMXB is minimum. Included in the 
spectral fitting discussed below is a term for the pulsar using the
best-fit to its {\it BeppoSAX} spectrum, with the normalization
reduced by the factor 0.725 appropriate to the offset position of the
X-ray pulsar in the PCA.

\subsection{{\it BeppoSAX}}

Data from the Medium-Energy Concentrator Spectrometer (MECS; 1.3--10
keV; Boella et al. 1997) on-board {\it BeppoSAX} are presented. The
MECS consists of three grazing incidence telescopes with imaging gas
scintillation proportional counters in their focal planes; however one
of these had failed prior to the observation of XB\th 1323-619. The MECS
is well-suited to the measurement of radial intensity profiles of a
source having an excess over the point spread function due to dust
scattering. The high energy
(HPGSPC and PDS) instruments are not useful since the scattering takes
place at low energies and the instruments are non-imaging.
XB\th 1323-619 was observed using {\it BeppoSAX} between 1997 August 22
17:06 and August 24 02:02 UTC (Ba\l uci\'nska-Church et al. 1999).
Data were selected having elevation above
the Earth's limb of $>$4$\degmark$ and were extracted using all
of the image. The exposure in the MECS was 70~ks. Background subtraction was
performed using standard files, but is not critical for this
relatively bright source.

\section{Results}

\subsection{The dust-scattering halo}

\begin{figure}[!ht]                    %Fig. 1
\begin{center}
\includegraphics[height=86mm,angle=270]{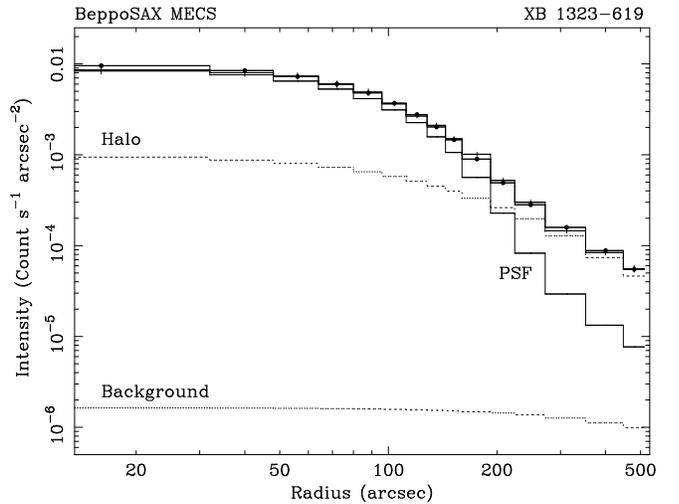} % - was [fig1]
\end{center}
\caption{Radial distribution of intensity of \src\ in the MECS in the
band 1.7--2.5 keV, together with the best-fit model consisting of halo, the point-spread
function and the background}
\label{}
\end{figure}

The MECS instrument allows accurate determination of the radial
dependence of intensity, and from measurements at
several energies, the dust-scattering cross section at 1~keV can be derived. 
The technique was based on that used by Predehl \& Schmitt (1995) in
their investigation of dust scattering in 25 Galactic sources using
the {\it Rosat} PSPC. The application of this technique to the {\it
BeppoSAX} MECS was developed and extensively tested in the case of
X\th 1624-490 (Ba\l uci\'nska-Church et al. 2000). We apply the same
technique here. The halo fraction $f_{\rm {h}}$ is defined {\it via}

\[f_{\rm {h}}\, =\, {I_{\rm {h}}\over I_{\rm {x}}\, +\, I_{\rm {h}}}\]

\noindent
where $I_{\rm {x}}$ is the observed source intensity and $I_{\rm {h}}$
the observed halo intensity. In the case of \src\ we find excesses in the radial 
distribution above the point spread function (PSF) for radii greater than 
$\sim $100$\arcsec$ revealing the halo. The radial profile was extracted using 
{\it XIMAGE}, radial bins were grouped to give a minimum of 20 counts per bin and 
systematic errors of 10\% were added between 10--100$\arcsec$ where the PSF is uncertain
by about this amount, and 2\% between 100--500$\arcsec$. The radial
profile was fitted between 0 and 500$\arcsec$, including contributions for the 
source convolved with the PSF, the halo, and also the background which was allowed to be a free 
parameter (as done by Predehl \& Schmitt). The halo was calculated on the basis of 
Rayleigh-Gans scattering theory (Predehl \& Klose 1996), although the halo fractions derived are
not strongly dependent on the model used (Predehl \& Klose 1996; Mathis
\& Lee 1991). The point spread function of Boella et al. (1997) was used.
\begin{figure*}[!ht]   %Fig. 2
\begin{center}
\includegraphics[width=60mm,height=160mm,angle=270]{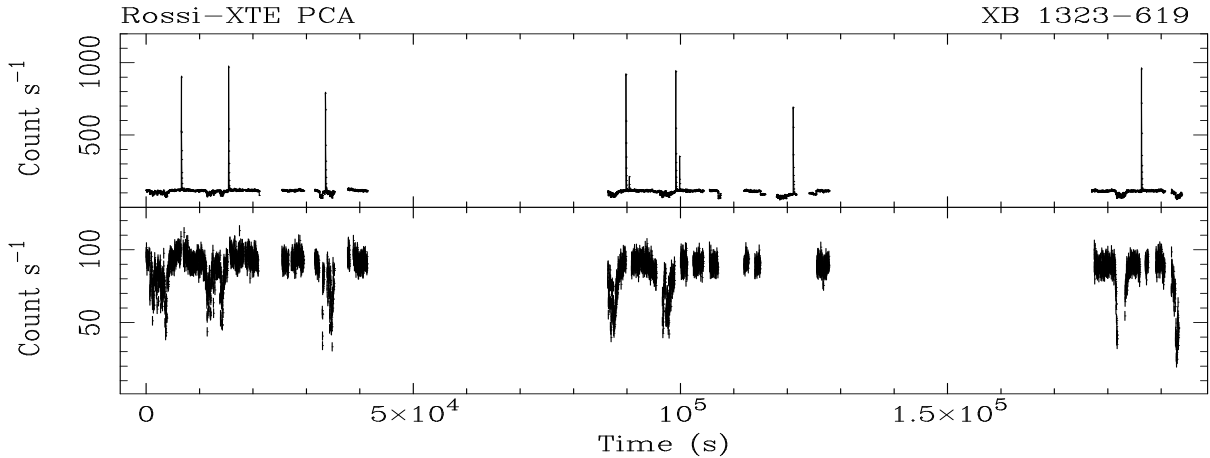}  %reverse x y - was [fig2_full]
\end{center}
\caption{Light curves of the 1997 {\it Rossi-XTE} observation of XB\th
1323-619 with 16~s binning in the band 2.0--20 keV.
The scales are chosen to show bursting clearly in the upper panel, and dipping clearly
in the lower panel. In both panels data were subjected to the standard screening
(Sect. 2.1); in addition, in the lower panel, data were selected as for spectral analysis
requiring all 5 PCUs to be operative; in the upper panel data are also included at
$\sim $120 ks when 1 PCU was switched off since this section of data contained one of the bursts}
\end{figure*}
For full details of the technique, see Ba\l uci\'nska-Church et al. 
(2000). Fits were made in several energy 
bands: 1.7--2.5, 2.5--3.5, 3.5--4.5, 4.5--5.6 and 5.5--6.5 keV,
for which values of $f_{\rm {h}}$ of 0.30$\pm $0.03,
0.18$\pm $0.02, 0.14$^{+0.01}_{-0.02}$,
0.12$\pm $0.05 and 0.07$^{+0.05}_{-0.06}$, respectively were found. 

The reduction in source intensity due to scattering out of the beam is given by $I_{\rm {x}}$ = 
$I_{\rm 0}\,{\rm e}^{-\tau}$, where $\tau $ is the optical depth to dust scattering, 
from which it follows that the halo fraction and $\tau$ are related by 
$f_{\rm h}$ = 1 - e$^{-\tau}$ provided the intensity scattered
out of the beam is balanced by that scattered into the beam (Martin 1970). It is 
known that the dust-scattering cross section varies approximately as $E^{-2}$ as
expected theoretically (Mauche \& Gorenstein 1986) and so $\tau $ will also vary 
as $E^{-2}$, so that

\[f_{\rm {h}}\, =\, {\rm 1}\,-\, {\rm e}^{-\tau_{\rm {1}} \, {\rm
E}^{\rm -2}}\]

\noindent
where $\tau_{\rm {1}}$ is the value at 1 keV. Thus from the halo fractions at 
several energies, we derived a value of the optical depth at 1 keV of 
$\tau $ = 1.8$\pm $0.4. This assumed a $E^{\rm -2}$ dependence which
appeared to fit the limited number of points adequately.

This value may be compared with the results of Predehl \& Schmitt (1995) who plot 
$\tau $ at 1 keV against $N_{\rm H}$ for the sources in their survey and derive a 
relation: $\tau $ = ${\rm 0.5}\,N_{\rm H}\,[{\rm 10}^{\rm 22}]$ - 0.083. From
the best-fit value of column density from analysis of the {\it BeppoSAX} observation 
(Ba\l uci\'nska-Church 1999), of $\rm {3.88\pm 0.16\times 10^{22}}$ atom $\rm {cm^{-2}}$ we 
derive $\tau $ = 1.86, in good agreement with the measured value. These results will 
be used in fitting the non-dip and dip {\it Rossi-XTE} data, and for determining the 
extent of the halo contribution to the {\it Rossi-XTE} 1.5--20 keV light curve (Fig. 2).

\subsection {The Rossi-XTE X-ray lightcurve}

Fig. 2 shows the 2.0--20 keV light curve of \src with 16~s binning. The scales 
are chosen to show dipping clearly in the lower panel, and bursting clearly 
in the upper panel. Parts of 7 dips can be seen, and 7 X-ray bursts,
two of which are double. 
In the lower panel, data were selected as for spectral analysis to have all 5 PCUs 
operative plus other screening as described in Sect. 2.1; however, there is a 
small part of the observation at $\sim $120 ks
when one PCU was not operative. In the upper panel these data are included 
which contain one of the bursts, for which the available count rate is 
consequently reduced by a factor of 0.8. This additional burst was included in 
analysis of the burst repetition  rate. The bursts were
found to repeat on a timescale of 2.45--2.59 hr, and the data gaps
during which other bursts most probably occurred are also consistent
with this recurrence timescale. In the previous {\it BeppoSAX}
observation of 1997, August 22, the recurrence timescale of bursting
was found to be 2.40--2.57 hr (Ba\l uci\'nska-Church et al. 1999).
This was significantly reduced compared with the timescales in the
{\it Exosat} observation (1985, February 13) of 5.30--5.43 hr (Parmar
et al. 1989) and in {\it ASCA} (1994, August 04) of 3.05 hr (Ba\l uci\'nska-Church
et al. 1999). The present observation was made on 1997, April 25, 5 months before the
{\it BeppoSAX} observation. If we assume, for example, that the rate of
change was constant between the {\it Exosat} and {\it BeppoSAX}
observations, the change per year is 0.23 hr, and the expected
difference between the {\it BeppoSAX} and {\it RXTE} values would be
only 0.1 hr. Further observations will reveal whether the recurrence
timescale continues to decrease, or shows any dependence on luminosity
(see discussion in Ba\l uci\'nska-Church et al. 1999).

In the dipping, considerable 
variability can be seen; some dips are deep and narrow while others 
are broad and shallower. The depth of the dipping is $\sim $60\%.
A trend of decreasing (non-dip) intensity by $\sim $5\% 
can be seen over the complete observation, and for this reason only the first 40 ks 
were included in spectral analysis. In Fig. 3, light curves in two energy bands 
2--4 keV and 4--20 keV are shown folded on the orbital period of 2.938 hr 
(Ba\l uci\'nska-Church et al. 1999), together with the hardness ratio formed from 
these. This clearly demonstrates the spectral hardening in dipping,
the strong variability in dipping in this source, and that the envelope of
dipping persists for a large fraction (40\%) of the orbital cycle.

\begin{figure}[!ht]   %Fig3
\begin{center}
\leavevmode
\includegraphics[width=63mm,angle=270]{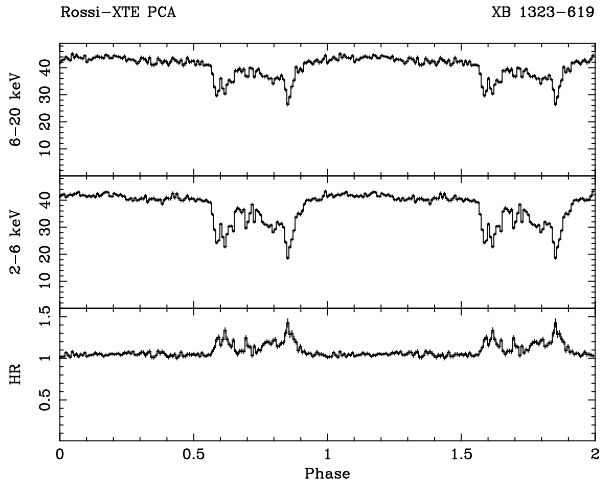}     % - was [fold]
\end{center}
\caption{Light curves in two energy bands folded on the orbital period
of 2.938 hr. The lower curve shows the hardness ratio formed by
dividing the light curves in the bands 4--20 keV and 2--4 keV}
\end{figure}

To demonstrate the very strong variability in
dipping, we show in Fig. 4 an expanded view of the 2.0--20 keV light curve 
at $\sim $30 ks from the start of the observations. In particular, it
can be seen that each dip consists of at least 5 individual
absorption events corresponding to individual blobs of absorber; in
addition, most of these events show further structure. This was found
to affect the results of spectral fitting in the case that dip spectra
were selected by making intensity slices. Clearly, blobbiness will
result in mixing data with different column densities, which will
be equivalent to mixing data at different intensity levels during
dipping. It is known that mixing intensity levels, e.g. by having
intensity bands too wide, causes problems in spectral fitting
(e.g. Church et al. 1998b). The point-source
blackbody emission component contained in the best-fit model will in
this case be a mixture of data in which the point-source is covered
by a blob and data in which it is not covered. The result will be seen as
unexpectedly low column densities for this component. Because of this,
data were also selected from a single dip labelled {\it A} in Fig. 4 
which appears to have little sub-structure. These spectra demonstrate
blackbody column densities substantially higher than for the
extended Comptonized emission as expected (see Sect. 3.3).

\begin{figure}[!ht]   %Fig4
\begin{center}
\leavevmode
\includegraphics[width=63mm,angle=270]{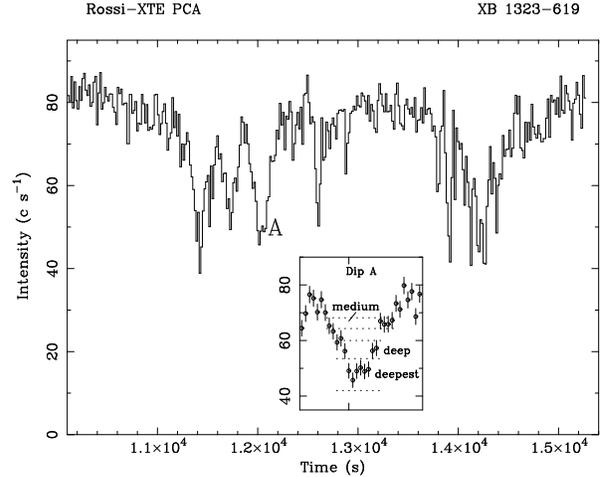}     % - was [fine]
\end{center}
\caption{Expanded light curve in the band 2.0--20 keV demonstrating
the strong variability during dipping and the blobby nature of the
absorber. The dip {\it A} is used (below) in the selection of dip
spectra; the inset shows the intensity bands used for the non-dip and
3 dip spectra}
\end{figure}

\subsection{Spectral Evolution During Dipping}

Two sets of spectra were used for spectral analysis. Firstly, data were
selected from the first 40 ks of the observation in intensity bands;
secondly, intensity selection was made using only the data from the
individual dip {\it A} in Fig. 4. In the first case, 
data from all 5 PCUs were selected 
corresponding to non-dip emission and four levels of dipping chosen 
using bands 85--87 c s$^{\rm -1}$ for non-dip, 60--65, 50--55, 40--45
c s$^{\rm -1}$ for intermediate dipping and 35--40 c s$^{\rm -1}$ for deep dipping.
These intensity bands were made relatively narrow to avoid as
much as possible mixing data from different intensities in each
spectrum. Care was also taken to exclude
all traces of bursting in selecting non-dip data. An intensity band
close to non-dip was avoided as experience has shown that model parameters
are not well-determined in very shallow dipping.
These spectra were used to eliminate various spectral models.
A second set of spectra were produced from the single
sub-dip {\it A}, consisting of a non-dip plus 3 dip levels.
These spectra were used for final fitting, allowing comparison of
results with results for the first set of spectra in which
dip data containing a high degree of variability in
dipping are superimposed.

Using data from the first 40 ks of the observation, simple models were tested
by fitting the non-dip spectrum, including absorbed bremmstrahlung
({\sc ab$\ast$br}), absorbed blackbody ({\sc ab$\ast$bb}), absorbed power law
({\sc ab$\ast$pl}) and cut-off power law ({\sc ab$\ast$cpl}). The one-component
thermal models can be rejected straight away, the blackbody model giving
$\chi^2$/dof = 2590/53, and the bremsstrahlung model giving 143/55. It is
clear that the non-dip spectrum by its approximately power law nature is dominated 
by Comptonization, and we consider these models no further. 
Next, simultaneous fitting of all 5 spectra was carried out in the band 2.5--25 keV 
where the source is significantly detected above the background, this constraining 
possible models much more strongly that fitting a single non-dip spectrum.
In each case, the emission
parameters specifying the model were chained to be equal for all 
spectra fitted, and only the absorption terms were allowed to vary.
Our best-fit model (below) requires inclusion of an Fe line, of the
pulsar contribution and the effects of the halo. Consequently, we 
performed a second stage of testing the other models in which the 
additional 3 terms were also included.
Results are shown in Table 1 for the {\sc ab$\ast$pl}, {\sc
ab$\ast$cpl} models, and for progressive covering of Comptonized emission 
model: {\sc pcf$\ast$cpl}. Finally, the two-component model
{\sc ab$\ast$bb + pcf$\ast$cpl} was tried, consisting of blackbody emission identified
with the neutron star plus Comptonized emission from an extended ADC.
The point-like blackbody is covered rapidly whereas the extended Comptonized
emission is progressively covered as the absorber moves across the source.
Comptonized emission was modelled by a cut-off power law as the energy
range of the PCA extends into the region where Comptonization
down-curving is expected, and a power law approximation would be
invalid. However, the cut-off energy $E_{\rm {CO}}$ is not very well constrained in
the PCA, and so this was fixed at the value obtained from the broad band
of {\it BeppoSAX} of 44 keV (Ba\l uci\'nska-Church et al. 1999).
%%\tabcolsep .8
\begin{table}[!ht]                 %Table 1
\caption{Results of fitting simultaneously the 5 PCA non-dip and dip spectra
selected from the first 40 ks of data
with various one-component and two-component models.
In each case the model also has an Fe line, a pulsar contribution and
halo terms as required in the case of the best-fit {\sc bb}+{\sc cpl}
model (see text). $N_{\rm H}$ is in units of $\rm {10^{22}}$ atom
cm$^{-2}$}
\begin{center}
\begin{tabular}{lccccr}
\hline\noalign{\smallskip}
Model &{\hskip -4mm}N$_{\rm H}^{\rm Gal}$ &{\hskip -4mm}$kT$ &{\hskip -4mm}$\Gamma$ &{\hskip -4mm}E$_{\rm CO}$
&{\hskip -3mm}$\chi^{\rm 2}$/dof\\
      &                      &{\hskip -4mm}keV  &         &{\hskip -4mm}keV\\
\hline\noalign{\smallskip}
{\sc pl}          &{\hskip -4mm} 4.7$\pm$0.1   &{\hskip -4mm} \dots         &{\hskip -4mm} 1.94$\pm$0.02 &{\hskip -4mm} \dots  &{\hskip -3mm}1086/266 \\
{\sc cpl}         &{\hskip -4mm} 4.5$\pm$0.1   &{\hskip -4mm} \dots         &{\hskip -4mm} 1.89$\pm$0.01 &{\hskip -4mm} $>$180 &{\hskip -3mm}1133/265 \\
{\sc pcf$\ast$cpl}&{\hskip -4mm} 6.1$\pm $0.3  &{\hskip -4mm} \dots         &{\hskip -4mm} 1.91$\pm$0.02 &{\hskip -4mm} $>$117 &{\hskip -3mm}336/261 \\
{\sc bb+pcf$\ast$cpl} &{\hskip -4mm} 2.7$\pm$0.6&{\hskip -4mm} 1.36$\pm$0.06 &{\hskip -4mm} 1.23$\pm$0.07 &{\hskip -4mm} 44    &{\hskip -3mm} 259/256 \\
\noalign{\smallskip}\hline
\end{tabular}
\end{center}
\end{table}
It can be seen that there is a significant improvement in fit by
adding the blackbody to the progressively covered Comptonized
component (third model). An F-test showed
that the presence of the additional term is significant at $>>$99.9\% 
confidence level. A similar result was obtained from the {\it BeppoSAX}
observation of \src\ (Ba\l uci\'nska-Church et al. 1999), so that the blackbody, 
although weak, is shown to be present. 

Use of the two-component model without a line revealed residuals at
about the position of an iron line, too strong to be due to remaining
uncertainty in the instrument response function, and so a Gaussian line
was added to the model. The non-dip spectrum provided a line energy of
6.43$\pm $0.21 keV; the line width $\sigma $ was fixed at an appropriate
value (0.4 keV) as is usually required to stabilise the fitting of a 
relatively broad line. This width was found to be approximately
correct, but free fitting of $\sigma $ was not possible as the value
tends to increase to several keV as part of the continuum becomes
incorrectly modelled by the line. The equivalent width of the line was found to be 
110$\pm $55 eV. An upper limit EW = 344 eV was found for a broad line 
using the {\it Exosat} GSPC (Gottwald et al. 1995). The rather short 
exposure of 16 ks with {\it ASCA} also allowed only an upper limit 
EW to be obtained, equal to 26 eV although this was for an assumed energy 
of 6.7 keV (Asai et al. 2000). In dipping, the line was modelled firstly 
as a component subject only to Galactic absorption, i.e. without additional 
absorption in dipping, but with free normalization. The results of this 
were not conclusive; although there was a decrease of line intensity in 
shallow dipping, the line appeared at about its non-dip strength in the 
deepest dip spectrum. This is further discussed below.

The two-component model used at this stage has the form:

\noindent
{\sc ag}$\,$e$^{-\tau} [\,${\sc ab}$\ast${\sc bb}+{\sc pcf}$\ast${\sc cpl}+{\sc
gau}]$\,$+{\sc ag}$^{'}$$\ast${\sc pl}{\vskip 0.mm\hskip 35mm} +{\sc
ag}$\,$[1$\,$-$\,$e$^{\tau}$]$\,$[{\sc bb}+{\sc cpl}+{\sc gau}]

\noindent 
Dust scattering is included by the
factor e$^{-\tau}$ which represents scattering out of the beam, and
(1 - e$^{-\tau}$) for scattering into the beam. Non-standard spectral
components were produced for these for use within the {\it XSPEC}
package. For the non-dip spectrum, the loss by scattering is balanced
by the intensity scattered into the beam. In dip spectra, the gain
depends on the non-dip intensity while the loss depends on the dip
intensity so that the gain exceeds the loss.
The X-ray pulsar is included as a constant power law term which does
not vary during dipping having its own Galactic column density 
{\sc ag$^{'}$} of ${\rm 7.8\times 10^{22}}$ atom cm$^{\rm -2}$ and power
law index 1.0 (Angelini et al. 1998). Results are given in Table 2.
In this fitting the line was included as a constant component; allowing
the normalization to vary did not substantially improve the quality of
the fits.
%%%\tabcolsep 8.0
\begin{table}[!h]                  %Table 2
\caption[ ]{Results of fitting non-dip and dip spectra from the first
40 ks of the observations with the
model containing continuum terms, an Fe line, the pulsar contribution
and the effects
of dust scattering both out of, and into, the beam. $N_{\rm H}$ is
given in units of $\rm {10^{22}}$ atom cm$^{-2}$}
\begin{center}
\begin{tabular}{lrrlr}
\hline\noalign{\smallskip}
Spectrum &$N_{\rm {H}}^{\rm {BB}}$&$N_{\rm {H}}^{\rm
{CPL}}$&$f$&$\chi^{\rm {2}}$/dof\\
\noalign{\smallskip\hrule\smallskip}
Non-dip &  2.7$\pm $0.1 &2.7$\pm $0.1       &0                 &61/51\\
Shallow &  9$\pm $1     &70$\pm $6          &0.45$\pm $0.01  &59/51\\
Medium  & 15$\pm $2     &79$\pm $6          &0.60$\pm $0.02  &58/51\\
Deep    & 28$\pm $4     &113$\pm $14        &0.69$\pm $0.02  &42/51\\
Deepest & 40$\pm $7     &179$^{+43}_{-30}$  &0.73$\pm $0.02  &29/46\\
\noalign{\smallskip}\hline
\end{tabular}
\end{center}
\end{table}

It can be seen that the covering fraction $f$ increases in a systematic
way as dipping gets deeper, with the column density $N_{\rm H}^{\rm CPL}$ 
increasing as the overlap between extended absorber and the extended ADC 
source becomes larger. However, the blackbody column density 
$N_{\rm H}^{\rm BB}$ is smaller at each level than the column of the extended 
emission component. In the present observation, the source is weak and 
the absorber very blobby so that selection in intensity bands {\it will} 
result in mixing data in which $N_{\rm H}$ for the blackbody varies between 
$\sim $zero and high values, as the line-of-sight to the point-source
passes through the blobby absorber, and so low values of $N_{\rm H}$
will be obtained.
Moreover, in our previous analyses of other dipping sources
(e.g. XB\th 1916-053, Church et al. 1997; X \th 1624-490, Smale et al.
2001), the blackbody has always had a high $N_{\rm H}$ consistent with 
being covered by the denser central regions of the absorber, whereas
the extended Comptonized emisssion has a lower $N_{\rm H}$ as it
\begin{figure*}[!ht]   %Fig. 5
\begin{center}
\leavevmode
\includegraphics[height=80mm,angle=270]{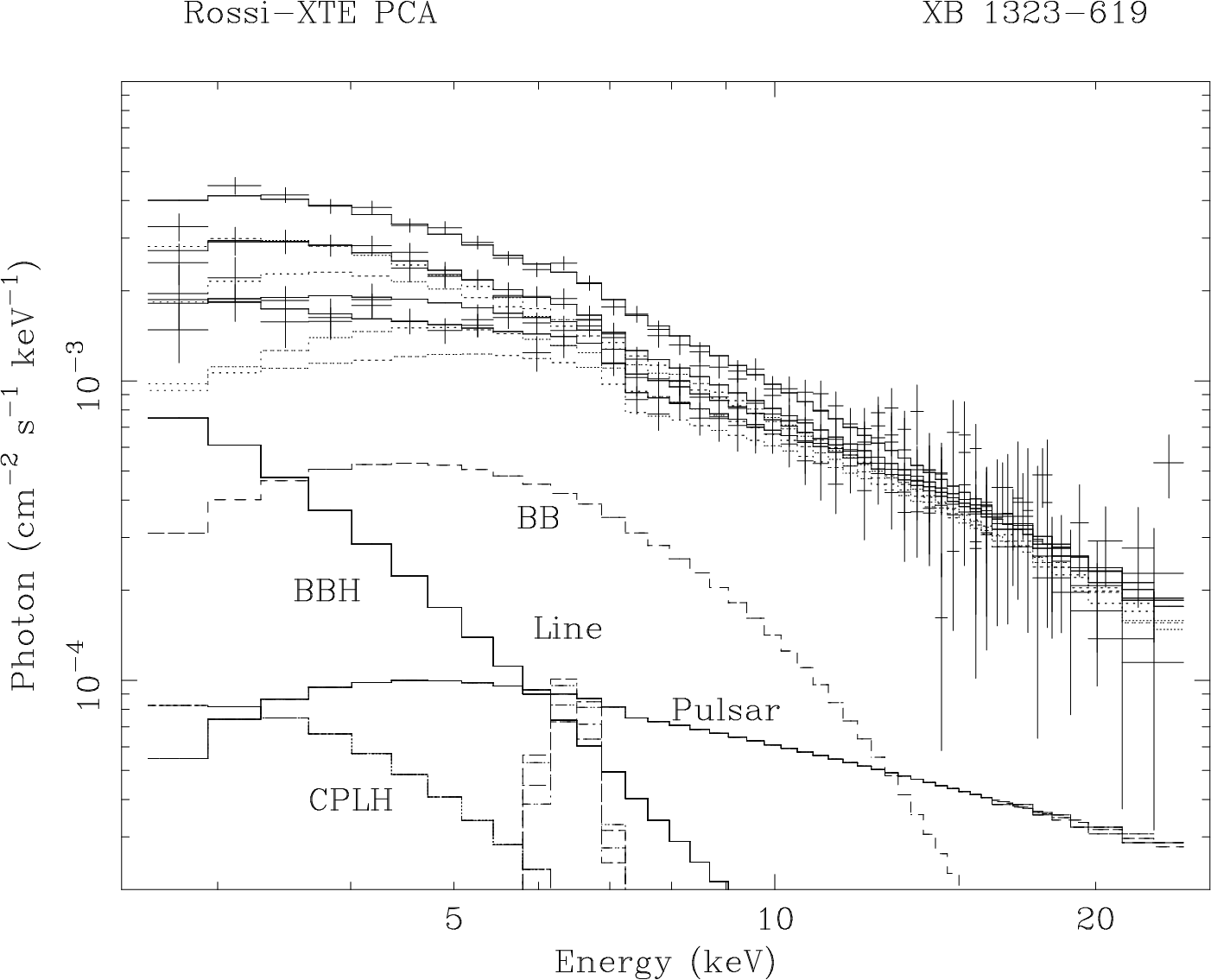}    % - was [better4p]
\includegraphics[height=80mm,angle=270]{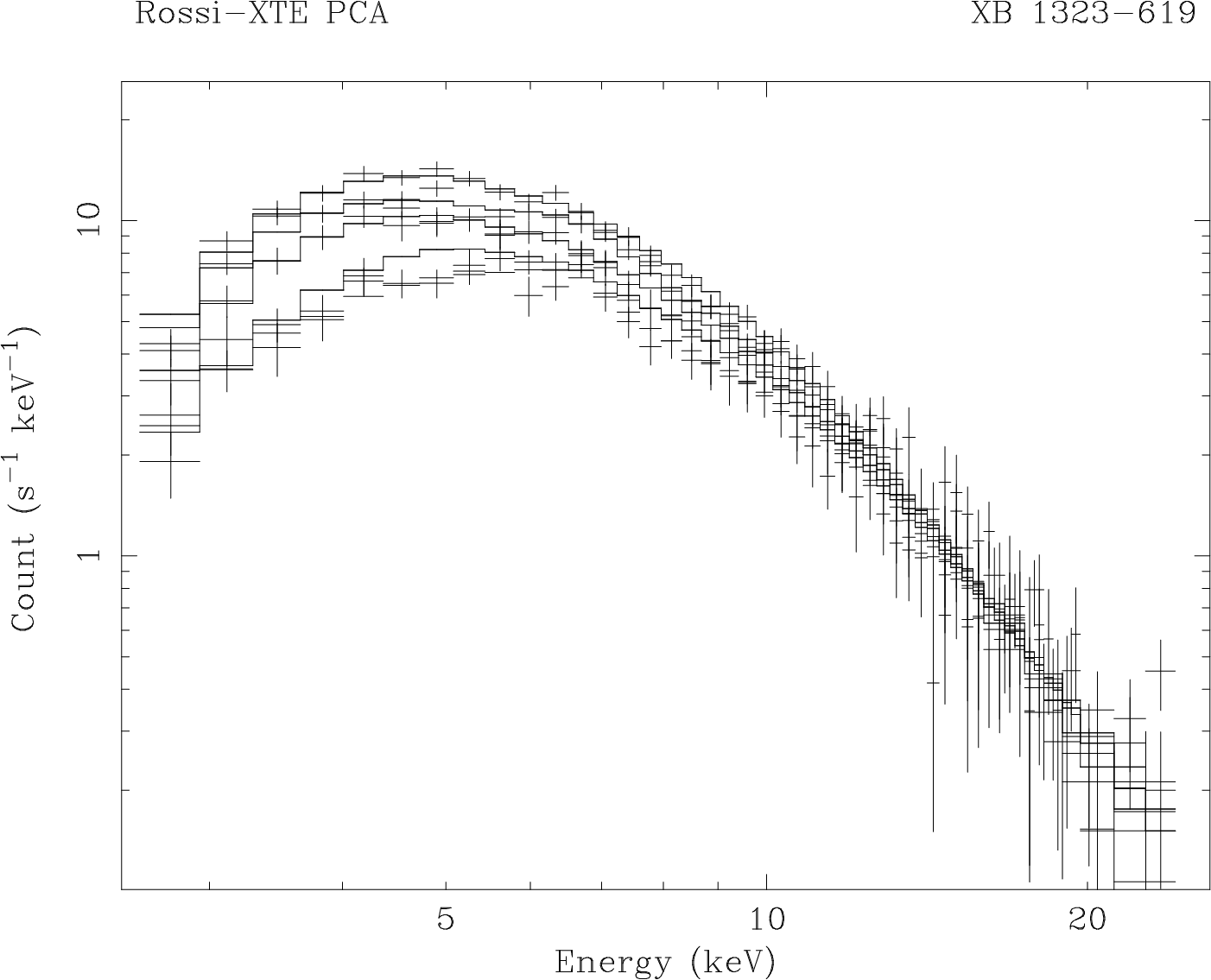}    % - was [better4c]
\end{center}
\caption{The four spectra from dip {\it A} fitted with the best-fit
model. Left: unfolded spectrum; Right: folded spectrum.
The model includes the Comptonized and total model terms {\it plus}
the blackbody (totally absorbed in dipping), the pulsar and line terms
as indicated and the cut-off power law halo {\sc cplh} and the
blackbody halo {\sc bbh}}
\end{figure*}
averages across the absorber.

To investigate these effects, we next analysed the non-dip and 3 dip
level spectra selected
from the single sub-dip {\it A}. Results are shown in Table 3 and
plotted in Fig. 5 as both unfolded and folded spectra.
%%%\tabcolsep 7.0
\begin{table}[!ht]                  %Table 3
\caption[ ]{Results of fitting non-dip and dip spectra from dip {\it A}
with the best-fit model. $N_{\rm H}$ is in units of $\rm {10^{22}}$
atom cm$^{\rm -2}$}
\begin{center}
\begin{tabular}{lrrlr}
\hline\noalign{\smallskip}
Spectrum &$N_{\rm {H}}^{\rm {BB}}$&$N_{\rm {H}}^{\rm
{CPL}}$&$f$&$\chi^{\rm {2}}$/dof\\
\noalign{\smallskip\hrule\smallskip}
Non-dip &  3.9                     &3.9            &0                 &42/42\\
Medium  & ${\rm 1.0\times 10^3}$   &7.7$\pm $1.3   &0.56$\pm $0.20  &30/38\\
Deep    & ${\rm >4.0\times 10^4}$  &20$\pm $4 &0.70$\pm $0.08  &37/42\\
Deepest & ${\rm >4.0\times 10^4}$  &34$\pm $4 &0.65$\pm $0.04  &26/38\\
\noalign{\smallskip}\hline
\end{tabular}
\end{center}
\end{table}

These results show that when data is selected more carefully using
only a single strong dip, good fits are obtained to all spectra, 
with substantially improved values of $\chi^2$/dof. 
The best-fit model consisted of a blackbody with $kT$ = 1.79$\pm $0.21 keV plus
an extended cut-off power law, having power law index $\Gamma $ =
1.61$\pm $0.04 keV, with the cut-off energy being fixed at the {\it BeppoSAX}
value of 44 keV. In dipping, the blackbody column density is larger than that of the extended
emission component as expected. The results are entirely consistent with
the two-component emission model consisting of point-like blackbody
emission from the neutron star and extended Comptonized emission from
the ADC. The results for the Fe line were also more conclusive in that
there was a definite decrease of line intensity in the shallow and
medium dip spectra, although, as before, the line became stronger in
deep dipping. Thus acceptable fits were obtained by including the
line in the same covering factor as for the extended Comptonized
emission, as shown in Fig. 5.

\section{Discussion}

The best-fit model results compare reasonably well with the values obtained
from {\it BeppoSAX} of kT = 1.77$\pm$0.25, and $\Gamma$ =
1.48$\pm$0.01. The 1--10 keV luminosity of \src\ was ${\rm 3.0\times 10^{36}}$ erg
s$^{\rm -1}$ in the {\it Rossi-XTE} observation, compared with
${\rm 1.9\times 10^{36}}$ erg s$^{\rm -1}$ during the {\it BeppoSAX}
observation. The combined effect of the halo and the X-ray pulsar
amounts to $\sim $8\% of the non-dip intensity. Thus, the observed
depth of dipping of $\sim $60\% would be increased to $\sim $70\% 
without these effects, consistent with the maximum covering fraction
determined in dipping. Dipping does not however, reach 100\% deep,
due mostly to the absorber having a blobby structure
allowing transmission of radiation between the blobs.

From this observation, we have detected an iron line in 
XB\th 1323$-$619. The energy of the line at 6.43$\pm$0.21 keV is 
interesting because of
the question of whether it can originate in the ADC as the evidence
indicates. It has
been known for some time that iron lines in LMXB tend to have energies
of $\sim $6.6 keV (Asai et al. 2000; White et al. 1985, 1986), 
suggesting origin in the ADC produced by photoionization followed by recombination. 
However, Smale et al. (1993) discussed the possible sites of 6.7 keV emission in 
Cyg\th X-2 and concluded that the disk was the origin of the emission.
In the {\it ASCA} iron line survey (Asai et al. 2000),
the mean energy for 20 sources was
6.56 keV, with only 5 sources having measured energies of 6.5 keV or lower. 
The mean energy implies a relatively low ionization state as it is
equivalent to an ionization parameter $\xi$ of $\sim$100, 
where $\xi$ is $L/n\,r^2$. Hirano et al. (1987)
had conducted a similar study of iron lines in LMXB using {\it Tenma}
and also carried out simulations of line emission from an ADC 
of varying ionization state. This showed that iron fluorescence is also
possible in the ADC for values of the ionization parameter $\sim$100.
Other authors have measured energies of $\sim $6.4 keV in particular
LMXB sources and have suggested that the line originates in the accretion disk
(e.g. Barret et al. 2000). In the dipping, flaring source X\th 1624-490, we have
also detected a broad iron line at 6.4 keV, this agreeing with the energy found by
Asai et al. (2000), and investigated the variation of the line in
dipping (Smale et al. 2001) and in flaring (Ba\l uci\'nska-Church et al. 2001).
In dipping, the line variation is well-described by giving it the same covering 
fraction as the extended Comptonized emission of the ADC, strongly suggesting 
that the line originates in the ADC, even though the energy may be
regarded as low for this. In flaring, it was found that the line intensity 
correlated strongly with the luminosity of the neutron star blackbody emission,
providing direct evidence that the line is excited by the central source.
Similarly, the 0.65 keV line in XBT\th 0748-676 (Church et al. 1997), 
has the same covering fraction as the
Comptonized emission. In the present case of XB\th 1323-619, the line  
varies in dipping in approximately the same way as the Comptonized emission,
except for the apparent presence of the line in deep dipping, which is
not understood. It is possible that the measurement is complicated by
the fragmented nature of the absorber. Further work will be necessary
to clarify the line behaviour in dipping and we will investigate this in
more detail using the observation of \src\ with {\it XMM} that is scheduled.
Thus the evidence supports 
origin of the line in the ADC as in other LMXB, although its energy
implies it is fluorescent. Many workers argue that it is not possible
to have a low ionization state in the ADC since the electron
temperature can be high ($kT_{\rm e}$ $\approxgt$ 15 keV in this
source; Sect. 1). However, the {\it ASCA} LMXB line survey
shows that the ionization state {\it is} relatively low, since the mean
energy of 6.56 keV corresponds to $\xi$ $\sim $100 (Dotani, priv comm)
so that there is a discrepancy between observation and simple theoretical expectations.

We also comment on the lack of detection of a reflection component in
the present data. In fact, there has been a general lack 
of detections of reflection components in LMXB, and reported
detections in a very small number of sources
have been ambiguous. In {\it Ginga} work on
LMXB, a broad edge-like structure above $\sim$7 keV in XB\th 1608-522 
was fitted either by partial absorption or a reflection component
(Yoshida et al. 1993). The {\it ASCA} survey of
LMXB (Church \& Ba\l uci\'nska-Church 2001) failed to detect
reflection in any of the sources investigated. We can argue (see also 
Ba\l uci\'nska-Church et al. 2001) that this may be due to the large size 
of the ADC, discussed below,
typically having radius $\sim $50,000 km. One consequence of this 
is that the accretion disk will be shielded from exposure to the neutron star 
source by the hot reflector of the ADC preventing reflection in the disk. 
Illumination of the disk
by the ADC will not produce an observable reflection component given
the large optical depth of the corona (Church 2001).
In black hole binaries, e.g. in 
Cyg\th X-1, the ADC appears to be much less extended
(Church 2001) so reflection may take place. 

Spectral fitting shows that the Comptonizing ADC region is clearly
extended as it can only be modelled by a progressive covering fraction,
as we have found in the other dipping sources, e.g. XB\th 1916-053
(Church et al. 1997) and X\th 1624-490 (Smale et al. 2001). This is
reinforced by measurements of ADC radius based on dip ingress time
determination in several dipping sources (Church 2001) showing that
$r_{\rm ADC}$ is typically 50,000 km or 15\% of the radius 
of the accretion disk. In XB\th 1323$-$619, $r_{\rm ADC}$ is 22,000--54,000 km
(reflecting the uncertainty in ingress time). The smallest value
obtained so far is in XBT\th 0748-676 where we obtain $r_{\rm ADC}$
= 8,500$\pm$3,000 km from the dip ingress time, and 3,500$\pm$1,500 km
from the eclipse ingress time, consistent within the errors.
The largest value is 53,000 km in the bright source
X\th 1624-490 (Church 2001), approximately half the radius of the
accretion disk. Thus, in all sources the ADC radius is many times
larger than the neutron star radius. In the case of XBT\th 0748-676, Bonnet-Bidaud
et al. (2001) obtained $r_{\rm ADC}$ $\approx$ 2,000 km from the {\it XMM}
observation. However, the light curve of this observation was interpreted by them
as flaring separated by low intensity intervals, whereas the {\it ASCA}
light curve (Church et al. 1998a) with a similar shape was proven
by spectral analysis to represent several intervals of dipping per
orbital cycle. Thus the ``flaring'' actually consisted of a temporary return from
dipping to the non-dip state. This may have led to the somewhat smaller
value of $r_{\rm ADC}$ deduced. Apart from the very large radius,
it can also be argued that the ADC is {\it thin}
(Smale et al. 2001). The extended size has several significant 
consequences as indicated in Sect. 1.
Firstly, it does not allow models in which Comptonization 
is localized to the neighbourhood of the neutron star. Secondly,
it is not expected that disk blackbody emission will be observed
since the ADC (of high optical depth; Church 2001) covers 
all of the X-ray emitting disk so that all disk blackbody radiation will be
Comptonized, explaining naturally the dominance of Comptonization in 
LMXB. The expectation that disk blackbody will not be observed
agrees with the results of Church \&
Ba\l uci\'nska-Church (2001) from a survey of LMXB with {\it ASCA} and
{\it BeppoSAX} who found that in the majority of cases,
the inner disk radius was substantially smaller than 10 km, the
neutron star radius. 

The large size of the ADC has implications for
different representations of Comptonization.
There has been an increasing use in recent years (Guainazzi et al.
1998; Oosterbroek et al. 2001) in
analysis of {\it BeppoSAX} data of the model {\sc diskbb + comptt}
i.e. disk blackbody plus Comptonization described by the
{\sc comptt} model in {\it XSPEC} based on the prescription of
Titarchuk (1994). It is sometimes claimed 
that {\sc comptt} is a better representation of Comptonization at low
energies than a cut-off power law, which could be inaccurate
because of a lack of seed photons below 1 keV
However, a major consequence of an ADC radius of 50,000 km
is that the spectrum of the accretion disk {\it under the ADC} is very
soft. Calculation of this spectrum using the temperature profile $T(r)$
from thin disk theory shows that the spectrum peaks between 0.001 and
0.1 keV for source luminosities between ${\rm 10^{36}}$ and $\rm
{10^{38}}$ erg s$^{\rm -1}$, providing a huge sea of very soft photons.
Thus, the cut-off power law is perfectly applicable. 

Moreover, in applications of the {\sc comptt} model, it is often the case
that the temperature of the seed photons is allowed to become large,
i.e. $\sim$1 keV (e.g. Guainazzi et al. 1998). In the model, the seed photons 
are described by the Wien 
approximation, i.e assuming that the temperature of the seed photons $kT_{\rm w}$ 
is much less than the X-ray energies in the spectrum: $kT_{\rm w}$ $<<$ $h\nu$.
For values of $kT_{\rm w}$ $\sim $1 keV derived from spectral fitting, this condition
is only satisfied above $\sim $5 keV so that use of this model leads to a 
substantial underestimation of the spectrum at energies below this. 

Finally, it should be pointed out that these arguments based on the large size of
the ADC provide a justification for the two-component model used here, and found
to fit the spectra of many LMXB. The model 
consisting of point-source blackbody emission from the
neutron star plus Comptonized emission from an extended ADC, seeded by the soft
photons from the disk vertically beneath the ADC, has implications
which are entirely consistent with observation. For example, if the ADC is large
and of high optical depth, it is not expected that diskblackbody
radiation will be observed, consistent with the {\it ASCA} survey results. On
the other hand, all accretion disk theory requires a substantial fraction of the total
emission to be from the neutron star. Modelling of neutron star
atmospheres (Madej 1991) shows that electron scattering does not
greatly modify the blackbody spectrum. Thus, blackbody emission
from the neutron star {\it should} be seen, and applying the
two-component model to many LMXB reveals that a blackbody component 
completely consistent with origin on the neutron star is
always present in varying degrees.

\begin{acknowledgements}

R.B. was funded by PPARC Grant 1997/S/S/02401

\end{acknowledgements}

\end{document}